\documentclass[%
 reprint,
 amsmath,amssymb,
 aps,
]{revtex4-2}

\usepackage{graphicx}
\usepackage{dcolumn}
\usepackage{bm}
\usepackage{hyperref}
\usepackage{braket}
\usepackage{color}
\usepackage{xcolor}
\usepackage[version=3]{mhchem}

\usepackage{hyperref} 
\hypersetup{
colorlinks=true,
    linkcolor=blue,     
    citecolor=blue,   
    urlcolor=blue
}

\usepackage{multirow}
\newcommand{\be}{\begin{equation}}
\newcommand{\ee}{\end{equation}}
\newcommand{\bea}{\begin{eqnarray}}
\newcommand{\eea}{\end{eqnarray}}
\newcommand{\Eq}[1]{Eq.~(\ref{#1})} 

\newcommand{\bos}[1]{\boldsymbol{#1}}
\newcommand{\mr}[1]{\mathrm{#1}}
\newcommand{\cm}{$\text{cm}^{-1}$}

\begin{document}

\preprint{APS/123-QED}

\title{Rotational Splittings in Diatomic Molecules of Interest to Searches for New Physics}

\author{Ayaki Sunaga}
\email{ayaki.sunaga@ttk.elte.hu, sunagaayaki@gmail.com}
\affiliation{ 
ELTE, E\"otv\"os Lor\'and University, Institute of Chemistry, P\'azm\'any P\'eter s\'et\'any 1/A 1117 Budapest, Hungary 
}%

\author{Timo Fleig}%
\email{timo.fleig@irsamc.ups-tlse.fr}
\affiliation{%
Laboratoire de Chimie et Physique Quantiques, FeRMI, Université de Toulouse, \\ 118 Route de Narbonne, F-31062 Toulouse, France
}%

\date{\today}

\begin{abstract}
Diatomic molecules with an energetically low-lying $^3 \Delta_1$ state are attractive platforms to detect new physics beyond the Standard Model, such as parity- and time-reversal violating phenomena. One of the advantages of using a $^3 \Delta_1$ state is its tiny $\Lambda$-splitting due to the coupling between the electronic and rotational angular momenta, which facilitates polarizing the molecules in small external electric fields. Theoretical estimation of the magnitude of the $\Lambda$-splitting is helpful for planning new experiments. In this study, we present a theoretical model to calculate the $\Lambda$-splitting. Our model integrates the relativistic four-component wavefunction and the traditional rotational Hamiltonian based on Hund's case (a). The multireference character of the wavefunction is taken into account. Our calculations for PtH and \ce{ThF+} molecules qualitatively agree with experiment. The $\Lambda$-splitting of \ce{TaO+} for the rotational ground state is predicted to be around 9 kHz. This tiny splitting can reduce the systematic uncertainty, but in a practical experiment, it may cause depolarization during rotation ramp-up.
\end{abstract}


\maketitle

\section{Introduction}
Diatomic molecules are used as powerful low-energy probes in the search for physics beyond the Standard Model (SM) of elementary particles \cite{https://doi.org/10.48550/arxiv.2203.08103}. In particular,
measurements and calculations on the hafnium flouride cation \ce{HfF+} \cite{JILA_eEDM_2023,PhysRevA.96.040502,Skripnikov_HfF+_JCP2017}
currently yield the strongest constraint on the electric dipole moment (EDM) of the electron~\cite{LeptonicCPviolation_RevModPhys2012,Shindler_EDMreview_2021}. Further advances are expected in the near future from work on the thorium monofluoride cation ThF$^+$~\cite{Ng2022PRA_ThF+_omega,Gresh_ThF+_JMS2016,Denis2015NJP,Skripnikov_ThF+_PRA2015,heaven_ThF+_JCP2012,Ng2025PRA_ThF+} and the tantalum monoxide cation TaO$^+$~\cite{PhysRevA.95.022504,Sunaga_Fleig_2022,PhysRevA.109.033107,Skripnikov_TaO+_2022} the latter of which can also be employed as a probe for nuclear charge-parity (CP) violation through the nuclear magnetic quadrupole moment~\cite{Flambaum2014PRL_MQM,Lackenby2018PRD_MQM}. 

In state-of-the-art experiments \cite{ACME_ThO_eEDM_nature2018,JILA_eEDM_2023} aiming to measure a molecular EDM in the laboratory frame it is required to mix opposite parity states by polarizing the molecule through an external electric field. This mixing depends on the separation of the molecular target rovibronic energy levels that is induced by the coupling of intrinsic angular momenta to the angular momentum of the molecule rotating in the laboratory frame. As an example, the electronic state in which the EDM measurement is carried out in the thorium monoxide (ThO) and the ThF$^+$ molecules is a $^3\Delta_1$ state where the total electronic angular momentum projection onto the internuclear axis is $\Omega=1$. This state exhibits \cite{ACME_2017_methods} quasi-degenerate pairs of rotational levels with well-defined parity and with energy splittings $\Delta_{\Omega}$ (or $\Delta_{\Lambda}$ where $\Lambda$ is the total electronic orbital angular momentum projection).
It is the purpose of this paper to present a method for calculating these so-called $\Omega$- or $\Lambda$-doublings and its application to molecules of interest in low-energy searches of CP-violation beyond that already known to exist in Nature~\cite{Kobayashi,LeptonicCPviolation_RevModPhys2012,Shindler_EDMreview_2021}.

In the following section \ref{SEC:THEO} we briefly discuss the theory underlying our approach and the specific approximations we make in view of the relevant experimental conditions. 
Next, we explain the mechanism in our model that leads to the $\Lambda$-doublet in the $^3\Delta_1$ state.
Typically, an EDM measurement is carried out in the rovibrational ground levels of either the electronic ground state or an energetically low-lying excited electronic state of the molecule~\cite{ACME_2017_methods,JILA_eEDM_2023}. Under these circumstances the required molecular vibrational overlap integrals can be approximated conveniently. In section \ref{SEC:APPL} we discuss applications of our approach. The initial application concerns the platinum monohydride (PtH) molecule. Rotational couplings have been calculated earlier and quite extensively for this molecule \cite{FleigM94,Fleig1996JMS} which allows us to draw comparisons and to verify that our present method is correctly implemented. We then go on to apply our approach to molecular ions that are being prepared to become leading contenders in EDM measurements, the ThF$^+$ and the TaO$^+$ molecular ions. 
We conclude on our findings in section \ref{SEC:CONCL}.

\section{Theory}
\label{SEC:THEO}
\subsection{$\Lambda$-doublet structure}
Earlier approaches to the calculation of molecular rotational couplings were based on a framework of scalar relativistic (or non-relativistic) wavefunctions and required the explicit calculation of matrix elements over the spin-orbit interaction Hamiltonian: the matrix elements were treated through perturbation theory
\cite{Brown1979JMS_Lambda,LambdaInDelta_JMS1987,deVivie1988MP_Lambda,Kozlov2009PRA_Lambda,Fan2025PRA_NEQ_doublet,Gordon2025JCP_Lambda_new} 
or matrix diagonalization \cite{Marian1995BBPC_NiH,Fleig1996JMS}.
   
The present theoretical approach closely follows the approach as described by Lef\`ebvre-Brion and Field \cite{brion} which uses an effective theory for many-body states in Born-Oppenheimer approximation represented in Hund's case (a) for diatomic molecules. The choice of a Hund's case (a) model is justified by earlier findings for the TaO$^+$ cation \cite{Sunaga_Fleig_2022} showing that molecular electronic states are represented to a very good approximation within this model. However, we use molecular electronic wavefunctions from a four-component Dirac-theory-based framework which includes the spin-orbit interaction more accurately and already in the zeroth-order wavefunctions.
Calculations of $\Lambda$-splittings including nuclear angular momenta have been reported \cite{Meerts1972JMS_NO,Meerts1976CP_HF_Lambda,Kozlov2009PRA_Lambda,Fan2025PRA_NEQ_doublet}. However, the hyperfine interaction presents only a minute perturbation that can be neglected given the other approximations made in the effective approach.

$\Lambda$-type doubling matrix elements have been determined for $^3\Delta$ molecular states by Brown {\it{et al.}} in 1987~\cite{LambdaInDelta_JMS1987}. Although our present theoretical formulation is quite different from that approach, the qualitative aspects of the coupling are the same.

The molecules' energy is represented by the molecular Hamiltonian
\be\label{eq:ele_rot_H}
\hat{H} = \hat{H}^\mr{ELE} + \hat{H}^{\text{ROT}},
\ee
where $\hat{H}^\mr{ELE}$ is an electronic Hamiltonian. For a rigid diatomic rotor the Hamiltonian representing the molecular rotational motion is
\begin{equation}
        \hat{H}^{\text{ROT}} = \frac{1}{2\mu R^2}\, {\hat{\bf{N}}^2}
        \label{EQ:HROTC0}
\end{equation}
where $\mu = \frac{m_1 m_2}{m_1 + m_2}$ is the reduced mass for the two fixed atomic nuclei with rest masses $m_1$ and
$m_2$, respectively, $R$ is the (constant) distance coordinate between the two nuclei and $\hat{\bf{N}}$ is the operator of rotational angular momentum. Its classical counterpart is angular momentum taken with respect to an origin lying in the center of mass of the diatomic molecule, and it is expressed in space-fixed (laboratory) coordinates.

The molecular rotational angular momentum operator $\hat{\bf{N}}$ can be represented in terms of electronic angular momentum operators as
\begin{equation}
        \hat{\bf{N}} = \hat{\bf{J}} - \hat{\bf{L}} - \hat{\bf{S}}
        \label{EQ:REXPRA}
\end{equation}
where $\hat{\bf{J}}$ is the vector operator of total angular momentum, $\hat{\bf{L}}$ of total electronic orbital angular momentum and $\hat{\bf{S}}$ of total electronic spin. 
Inserting Eq. (\ref{EQ:REXPRA}) into Eq. (\ref{EQ:HROTC0}), exploiting the fact that components of different angular momentum operators commute and straightforward manipulations yield the rotational Hamiltonian in Hund's case (a):

\begin{eqnarray}\label{eq:rot_ham_expand}
 \nonumber
        \hat{H}^{\text{ROT}} &=& \frac{1}{2\mu R^2}\, \left[ \hat{\bf{J}}^2 - \hat{{J}}_z^2 + \hat{\bf{L}}^2 - \hat{{L}}_z^2
            + \hat{\bf{S}}^2 - \hat{{S}}_z^2  \right] \\
  \nonumber
  \hspace*{1.0cm} &&
            - \left( \hat{{J}}^+ \hat{{L}}^- + \hat{{J}}^- \hat{{L}}^+ + \hat{{J}}^+ \hat{{S}}^-
                    + \hat{{J}}^- \hat{{S}}^+ \right) \\
  \hspace*{1.0cm} && \left.      + \hat{{L}}^+ \hat{{S}}^- + \hat{{L}}^- \hat{{S}}^+ \right]
        \label{EQ:HROTA0}
\end{eqnarray}

For the matrix representation of this operator we use explicit signed basis states
defined as (see \cite{brion}, p.221 ff.)
\begin{eqnarray}\label{eq:e_f_basis}
        \nonumber
        \left| e_{J \Omega \Lambda \Sigma} \right> &=&
        \frac{1}{\sqrt{2}} \left[ \left| J\;\; \Omega\;\; \Lambda\;\; \Sigma \right> + \left| J\;\; -\Omega\;\; -\Lambda\;\; -\Sigma \right> \right] \\
        \left| f_{J \Omega \Lambda \Sigma} \right> &=&
        \frac{1}{\sqrt{2}} \left[ \left| J\;\; \Omega\;\; \Lambda\;\; \Sigma \right> - \left| J\;\; -\Omega\;\; -\Lambda\;\; -\Sigma \right> \right]
        \label{EQ:ROT_EF_BASIS}
\end{eqnarray}
where $\Omega$, $\Lambda$ and $\Sigma$ are the projection quantum numbers of  $\hat{{J}}$, $\hat{{L}}$ and $\hat{{S}}$, respectively, onto the molecular axis and $J$ is the quantum number of the total angular momentum. Here we do not explicitly show another quantum number, parity, which can be obtained with $J$ for each $e/f$ state~\cite{Brown1975JPS_ef_parity}. 

In practice, 
the evaluation of corresponding matrix elements requires the expansion of the operators  $\hat{{J}}^+$ and $\hat{{J}}^-$ with molecule-fixed commutation rules in terms of operators with anomalous (space-fixed) commutation rules.
It is found (\cite{brion}, p.76) that
\begin{equation}
        \hat{{J}}^+ = \hat{{J}}_Z \alpha_Z^+ + \frac{1}{2} \left( \hat{{J}}_+ \alpha_-^+ + \hat{{J}}_- \alpha_+^+ \right)
\end{equation}
where $\hat{{J}}_Z, \hat{{J}}_+,$ and $\hat{{J}}_-$ act in space-fixed coordinates and $\alpha_I^j = {\bf{e}}_I \cdot {\bf{e}}_j$ are the
{\it{direction-cosine}} matrix elements with $I$ a space-fixed and $j$ a molecule-fixed coordinate and ${\bf{e}}$ specifies
a unit vector. In such a representation
$\hat{{J}}^+$ can be evaluated in a basis of states labeled as $\left| J\;\; M\;\; \Omega \right>$ where $J$ and $M$ are
the space-fixed total angular momentum and total angular momentum projection quantum numbers, respectively. The matrix element is expressed by
(\cite{brion}, p.78)
\begin{equation}
        \left< J\;\; M\;\; \Omega\pm 1 \right| \hat{{J}}^{\mp} \left| J\;\; M\;\; \Omega \right>
        = \hbar \left[ J(J+1) - \Omega(\Omega\pm 1) \right]^{1/2}.
        \label{EQ:JPM_OMEGA}
\end{equation}
We use this expression in the explicit evaluation of our matrix elements in Hund's case (a) formalism.

Even if the quantum numbers $\Lambda$ and $\Sigma$ are known to sufficient accuracy for a given state, the evaluation of the matrix elements of the Hamiltonian in Eq.~(\ref{EQ:HROTA0}) using the basis states in Eq.~(\ref{EQ:ROT_EF_BASIS}) requires the knowledge of the total orbital angular-momentum quantum number $L$ when the expression $\hat{L}^{\pm}\, \left| J\;\; \Omega\;\; \Lambda\;\; \Sigma \right>$, for example, needs to be calculated. However, $L$ is not an exact quantum number in a (relativistic) molecular field, and thus the value of $L$ in the respective ``$L$ complex'' \cite{Hougen1970_rot_ene_lev} used for the evaluation is always approximate. In the PtH molecule $L$ is rather well defined, but in ThF$^+$ and TaO$^+$ the situation is more ambiguous. We discuss these cases and our reasoned choices in the applications section below.

In our current approach vibrational degrees of freedom are treated as follows. 
In the framework of the Born-Oppenheimer approximation~\cite{born_opp} the molecular wavefunction $\psi_{\text{mol}}$ is separated 
\cite{Bransden-Joachain} 
into an electronic part $\psi_{\text{el}}$ and a vibrational part $\psi_{\text{vib}}$, 
\begin{equation}
    \psi_{\text{mol}} = \psi_{\text{vib}}({\bf{R}}) \psi_{\text{el}}({\bf{r}}_1, \ldots, {\bf{r}}_n;{\bf{R}})
\end{equation}
for an $n$-electron diatomic molecule where the electronic wavefunction depends parametrically on the nuclear positions.  
The basis functions given in Eq. (\ref{EQ:ROT_EF_BASIS}) purely describe electronic degrees of freedom and we denote these as $A$ in the corresponding bra and ket vectors. The vibrational wavefunction is given in terms of nuclear degrees of freedom and will be denoted by $v$. As an example for a given matrix element, we take one term of the rotational Hamiltonian representing the spin-uncoupling and write its matrix element as 
\begin{eqnarray}
 \nonumber
        && \left< A \; v \right| -\frac{1}{2\mu R^2}\, \hat{{J}}^+ \hat{{S}}^- \left| A' \; v' \right> \\
        &=&
        -\left< v \right| \frac{1}{2\mu R^2} \left| v' \right>
        \left< A \right| \hat{{J}}^+ \hat{{S}}^- \left| A' \right>
        \label{Sample_ME}
\end{eqnarray}
with
\be
|A\rangle,\left|A^{\prime}\right\rangle 
\in
\{\left| e_{J \Omega \Lambda \Sigma} \right>,\left| f_{J \Omega \Lambda \Sigma} \right>\},
\ee
where the first factor on the rhs. of Eq. (\ref{Sample_ME}) is an integral over nuclear coordinates and the second factor is an electronic integral. Although we do not explicitly write it in the equation, the electronic energy associated with $\hat{H}^\mr{ELE}$ is added to the diagonal part of the matrix elements.

Supposing that the potential-energy curves of the respective electronic states are sufficiently parallel near the equilibrium internuclear distance of the diatomic molecule, the overlap of the corresponding ground-state vibrational wavefunctions can be approximated as $\left< v | v' \right> \approx 1$ from which
\begin{equation}
        \left< v \right| \frac{1}{2\mu R^2} \left| v' \right> 
        = \frac{1}{2\mu R^2} \left< v | v' \right>
        \approx \frac{B(v)}{\hbar^2}
        \label{EQ:VIB_OVERLAP}
\end{equation}
where $B(v)$ is the rotational constant of the target electronic state. 
The uncertainty due to this approximation would be small (ca. 10\%) in the case of our target molecules, because the change of the spectroscopic constants via the transition between the dominantly-coupled states are small amounts~\cite{heaven_ThF+_JCP2012,PhysRevA.95.022504,Irikura2023JCP}.

Approximating the latter by the equilibrium rotational constant $B_\mr{e}$ (or by $B_0$ for the vibrational ground state, if available) the sample matrix element in Eq. (\ref{Sample_ME}) becomes
\begin{equation}
        \left< A\; v \right| -\frac{1}{2\mu R^2}\, \hat{{J}}^+ \hat{{S}}^- \left| A'\; v' \right> \approx
        -\frac{B_e}{\hbar^2}\, \left< A \right| \hat{{J}}^+ \hat{{S}}^- \left| A' \right>
        \label{EQ:S_UNCOUPL_DELTA}
\end{equation}
The current approach is justified for the purposes mentioned in the introduction.

We formulate the $\Lambda$-splitting based on Hund's case (a) above, but our electronic Hamiltonian includes the spin-orbit interaction, and thus $\Lambda$ and $\Sigma$ are not exactly good quantum numbers for our wavefunction. In relativistic wavefunctions, the signed $e$ and $f$ basis states defined in Eq.~(\ref{eq:e_f_basis}) can be generalized as follows
\be\label{eq:e_f_lincomb}
\left| e/f_{J \Omega \Lambda \Sigma} \right>\rightarrow \sum_i C_i\left|e/f_{J \Omega \Lambda_i \Sigma_i}\right\rangle.
\ee
The $e$ and $f$ bases are orthogonalized, and the linear combination coefficients $C_i$ are determined to satisfy the normalization condition. This linear expansion of the basis is in our model key to the description of the $\Lambda$-splitting in $^3\Delta_1$ states, as shown in later sections. 

In addition, for \ce{ThF+} and \ce{TaO+} molecules, we take the electronic configuration of each $\left|e/f_{J \Omega \Lambda_i \Sigma_i}\right\rangle$ basis into account because of the multireference character of these wavefunctions. This is not required for the PtH molecule since here the interaction space is comprised by one electronic configuration only.
The coupling occurs only when the electronic configurations (i.e., spinor structures) of bra and ket Hilbert-space vectors are the same, and the weight of the target electronic configuration has to be included as a factor. The target electronic configurations are $7s^2$ and $7s7p$ in the case of \ce{ThF+}, and $5d^2$ in the case of \ce{TaO+}.
When we denote the weight by the coefficient $d$,
the matrix element of the spin-uncoupling term in the $e$ basis and $f$ basis can be expressed by
\be\label{eq:d_appear}
\sum_{ij} C_iC'_j\, \left[d_i d'_j + \delta_{ij} (1-d_i d'_j) \right]\, \langle e_{J \Omega \Lambda_i \Sigma_i}| \hat{J}^{+} \hat{S}^{-}\left| f_{J' \Omega' \Lambda'_j \Sigma'_j}\right\rangle.
\ee
In the case of $i=j$, we do not multiply $d$ so that $\ket{e/f_{J \Omega \Lambda_i \Sigma_i}}$ can satisfy the normalization condition. The expression for the $i \neq j$ case indicates that only the target electronic configuration contributes to the matrix element.
Examples of the coefficients $C$ and $d$ are described in Secs.~\ref{SUBSEC:ThF+} and~\ref{SUBSEC:TaO+}.

\subsection{Mechanism of $\Lambda$-doubling in $^3 \Delta_1$ states}
In the following, we refer to vectors with exact quantum numbers in the $\Lambda\text{-}S$ picture as ``basis vectors'' or ``basis functions'' and those with approximate quantum numbers simply as ``states'', in order to avoid confusion.

The $e$-$f$ splitting does not occur in the $^3\Delta_1$ state in the lowest approximation that considers only the terms in \Eq{eq:rot_ham_expand} because neither $J^{\pm}L^{\mp}$ ($\Delta\Omega=\Delta\Lambda=\pm 1,\;\Delta\Sigma=0$) nor $J^{\pm}S^{\mp}$ ($\Delta\Omega=\Delta\Sigma=\pm 1,\;\Delta\Lambda=0$) operators can couple $\ket{\Lambda=2,\Sigma=-1}$ and $\ket{\Lambda=-2,\Sigma=1}$ basis functions~\cite{LambdaInDelta_JMS1987}.

Furthermore, there is also no direct rotational coupling between $^3\Delta_1$ and other electronic states. Any $\Lambda$-splitting must, therefore, be due to rotational couplings between excited electronic states that can mix with the $^3\Delta_1$ target state through a different mechanism.

We first use a simple model, assuming that the $e$-$f$ splitting of the $^3\Delta_1$ state occurs due to the contribution from a nearby $^{2S+1}\Pi_1$ term which rotationally couples with another energetically close 
$^{2S+1}\Sigma_0$ state. The contribution from the term $^{2S+1}\Pi_1$ is due to spin-orbit coupling and can be written as
\be\label{eq:lin_comb}
\ket{\Omega=1,\tilde{\Lambda},\tilde{\Sigma}} = a\ket{^3\Delta_1} + b\ket{^{2S+1}\Pi_1}.
\ee
$a$ and $b$ correspond to $C_{^3\Delta_1}$ and $C_{^{2S+1}\Pi_1}$ defined in \Eq{eq:e_f_lincomb}. The lhs. of Eq. (\ref{eq:lin_comb}) is to be understood as a physical state and the terms on the rhs. of Eq. (\ref{eq:lin_comb}) are basis functions.
Furthermore, $\tilde{\Lambda}=\braket{\Psi_{\Omega}|\hat{L}_z|\Psi_{\Omega}}$ and $\tilde{\Sigma}=\braket{\Psi_{\Omega}|\hat{S}_z|\Psi_{\Omega}}$ are now approximate quantum numbers that are close to integer/half-integer values. 
From the deviation between $\tilde{\Lambda}$ and $\Lambda$ ($\tilde{\Sigma}$ and $\Sigma$), we can assume the contribution from another basis function with a different value of $\Lambda$ ($\Sigma$). When $|\tilde{\Lambda}| < \Lambda=2$, a $\Pi_1$ basis function contributes to the $^3\Delta_1$ state, as shown in \Eq{eq:lin_comb}.

The coefficients $a$ and $b$ can be obtained by solving the simultaneous equations
\bea
2\times a^2 + (1)\times b^2 &=& \tilde{\Lambda}, \\ \nonumber
-1\times a^2 + (0)\times b^2 &=& \tilde{\Sigma}.
\eea
The normalization condition ($a^2+b^2=1$) is automatically satisfied in our model, as follows:
\begin{align}\label{eq:normalization}
\Lambda_a a^2 + \Lambda_b b^2  &= \tilde{\Lambda} \\ \nonumber
(\Omega -\Lambda_a) a^2 + (\Omega -\Lambda_b) b^2   &= \Omega - \tilde{\Lambda},
\end{align}
where $\Lambda_i$ corresponds to the $\Lambda$ value associated with the state of the corresponding coefficient ($i=a,b$). For example, in the case of the $^3\Delta_1$ state of \ce{ThF+}, $a=0.9911$ and $b=0.1334$ were obtained from $\tilde{\Lambda}=1.9822$ and $\tilde{\Sigma}=-0.9822$.

Although the magnitudes of $a$ and $b$ are determined without arbitrariness, the following approximations are included in this model: (i) We can consider the contribution from one basis vector ($\Pi_1$), and cannot determine the contributions from other non-$\Pi$ basis vectors. (ii) We cannot determine the (relative) sign of the linear combination coefficients ($a$ and $b$). The sign would not always be positive, as shown in the case of PtH (Table~\ref{tbl:Energy_PtH}). 
However, it does not affect the level ordering or the splitting size when the coupling between the two states is dominant, see Appendix~\ref{app:sign}.
(iii) Since the model takes only $\Lambda$ and $\Sigma$ into account, the spin multiplicity of $\Pi_1$ state is arbitrary. We selected the spin states that are energetically closest to the lowest-energy $^3\Delta_1$ state: $S=0$ for \ce{ThF+} and $S=1$ for \ce{TaO+}.

The $e$-$f$ splitting of the $^3\Delta_1$ state is due to a small contribution from the $^{2S+1}\Pi_1$ basis function that causes the $\Lambda$-doubling due to the coupling between the $^{2S+1}\Sigma_{0}$ basis functions. As an example, we show the analytical expression for the coupling between the $^3\Sigma_0$ and $^3\Pi_1$ basis functions:

\begin{widetext}
\bea\label{eq:3Sigm0_3Pi1}
\left\langle e_{3_{\Sigma_{0}}}\left|\left(\hat{J}^{+} \hat{L}^{-}+\hat{J}^{-} \hat{L}^{+}\right)\right|e_{3_{\Pi_{1}}} \right\rangle
&=&
\left\langle J \; 0 \; 0 \;0\left|\hat{J}^{+} \hat{L}^{-}\right|J \; 1 \; 1 \;0\right\rangle 
+\left\langle J \; 0 \; 0 \;0 \right|\hat{J}^{-} \hat{L}^{+}\left|J \; -1 \; -1 \;0\right\rangle \\ \nonumber 
&=& 
 \hbar \left[ J\left(J+1 \right)-1(1-1) \right]^{1/2} \hbar \left[ L\left(L+1 \right)-0(0+1) \right]^{1/2} \\ \nonumber 
&+& \hbar \left[ J\left(J+1 \right)-1(1-1) \right]^{1/2} \hbar \left[ L\left(L+1 \right)-0(0+1) \right]^{1/2} \\ \nonumber 
&=& 
\hbar^2 \left[ J\left(J+1 \right) \right]^{1/2} \left[ L\left(L+1 \right) \right]^{1/2}. 
\eea
\end{widetext}

The ordering of the quantum numbers is defined in Eq.~(\ref{eq:e_f_basis}).
To obtain the matrix element shown in \Eq{eq:d_appear}, the coefficients ($C$ and $d$) need to be determined.   
In the case of the coupling between the $^3\Sigma_{0}$ state and the $^3\Pi_{1}$ basis functions of the ground $^3\Delta_{1}$ state of \ce{TaO+} (cf. Table~\ref{tbl:TaO+_lin_comb}), 
\bea
C_{3_{\Sigma_{0}}}=1.0 &;& \quad C_{3_{\Pi_{1}}}=0.0283, \\ \nonumber
d_{3_{\Sigma_{0}}}=1.0 &;& \quad d_{3_{\Pi_{1}}}=0.43.
\eea
The other coupling terms are zero because of the cancellation
\bea
\left\langle f_{3_{\Sigma_{0}}}\left|\left(\hat{J}^{+} \hat{L}^{-}+\hat{J}^{-} \hat{L}^{+}\right)\right|f_{3_{\Pi_{1}}} \right\rangle&=&0, \\ \nonumber
\left\langle e_{3_{\Sigma_{0}}}\left|\left(\hat{J}^{+} \hat{L}^{-}+\hat{J}^{-} \hat{L}^{+}\right)\right|f_{3_{\Pi_{1}}} \right\rangle&=&0, \\ \nonumber
\left\langle f_{3_{\Sigma_{0}}}\left|\left(\hat{J}^{+} \hat{L}^{-}+\hat{J}^{-} \hat{L}^{+}\right)\right|e_{3_{\Pi_{1}}} \right\rangle&=&0 .
\eea
Similarly we can show that the S-uncoupling terms do not contribute to the splitting: the matrix elements between the $^3\Pi_{0/2}$ and $^3\Pi_{1}$ basis functions satisfy $\braket{e|(\hat{J}^{+} \hat{S}^{-}+\hat{J}^{-} \hat{S}^{+})|e}=\braket{f|(\hat{J}^{+} \hat{S}^{-}+\hat{J}^{-} \hat{S}^{+})|f}$ and $\braket{e/f|(\hat{J}^{+} \hat{S}^{-}+\hat{J}^{-} \hat{S}^{+})|f/e}=0$.

The coefficient for the electronic configuration $d$ defined in~\Eq{eq:d_appear} is obtained from the ratio of the target electronic configurations. 
For example, the $d$ of the ground $^1\Sigma_0$ of \ce{ThF+} is obtained from $d=0.75/(0.75+0.12)\approx0.87$, where 0.75 and 0.12 are the squares of the respective expansion coefficients of the $\left(7 s_{\sigma, 1 / 2}\right)^2$ and $\left(6 d_{\delta, 3 / 2}\right)^2$ Slater determinants (cf. Table~\ref{tbl:ThF+_lin_comb}). The value of $d$ of the correction basis function of the $^3\Delta_1$ state, $^1\Pi_1$, is obtained from the closest $^{1,3}\Pi_1$ state with the energy of 6639~\cm, where $d=(0.41+0.18)/(0.41+0.18+0.16)\approx0.78$. $d$ of \ce{TaO+} is obtained in the same manner.

\section{Application}
\label{SEC:APPL}
We applied the developed code to three molecules. The application to PtH, for which experimental values have been reported, largely serves for verification purposes of the present method. \ce{ThF+} is an important molecule in its own right and an example for showing the ambiguity of the quantum number $L$. Also here we can compare with experimental results. For \ce{TaO+} we are then able to make confident predictions for the expected $\Lambda$-splitting. The employed input parameters (electronic energy and rotational constants) and configuration coefficients ($C$ and $d$) are summarised in Appendix~\ref{app:data}.
The uncertainties of the calculations are roughly obtained from the input parameters, such as rotational constant, excitation energy, and $L$.

%
%
\subsection{Platinum hydride}
The $\Lambda$-splittings of the five lowest-energy electronic states of PtH are listed Table~\ref{tbl:splitting_PtH}. The difference between the labels TW-X (X = A to E) is the employed energy and rotational constants, which are listed in the table header.
These five states arise from the two atomic states, Pt's $^1D_2$ and H's $^2S_{1/2}$~\cite{note_mol_term}. Our theory and code successfully reproduce the order of magnitude for the available experimental values, but some input-parameter dependence is observed.

The input-parameter dependence of the $\Lambda$-splitting becomes significant when the relative difference of the energy gaps between the coupled electronic states is changed. For the $\Omega=3/2$ (I) state, TW-B shows the lowest splitting, while TW-D shows approximately twice the splitting. This is due to the employed energy difference between the $\Omega=3/2$~(I) and $\Omega=1/2$~(I) state: The energy difference of the former (latter) is 1~935~\cm\ (1~213~\cm), the relative difference of which is approximately 1.6.  
From the comparison between TW-A and TW-B, the contributions from the $\Omega=3/2$ (II), 1/2 (II) states to the $\Omega=3/2$ (I) state reach more than 10\% at small $J$, and go beyond 20\% at large $J$, even though its energy gap is ca. 8~000 \cm. 
When the experimental energy is used (TW-D and TW-E), $\Lambda$-splitting is overestimated. The calculated values might become closer to the experiment if the experimental energy were available for the $\Omega=1/2$ (II) state, and the energy difference from the $\Omega=3/2$ (II) state were increased. In the calculation of TW-D and TW-E, the employed energy difference between $\Omega=3/2$ (II) and $\Omega=1/2$ (II) states (350~\cm) is much smaller than the other conditions, 583~\cm\ (TW-A and TW-B) and 684~\cm\ (TW-C).  
The uncertainties (\%) for the $\Lambda$-splitting for the five electronic states at the lowest rotational states are evaluated from the mean values and their standard deviations. For states TW-B through TW-E, the obtained uncertainties are 32, 3, 32, 53, and 3 for $\Omega=$ 5/2, 1/2(I), 3/2(I), 3/2(II), and 1/2(II) states, respectively.

\begin{figure*}[htbp!]
    \centering
    \includegraphics[width=0.8\linewidth]{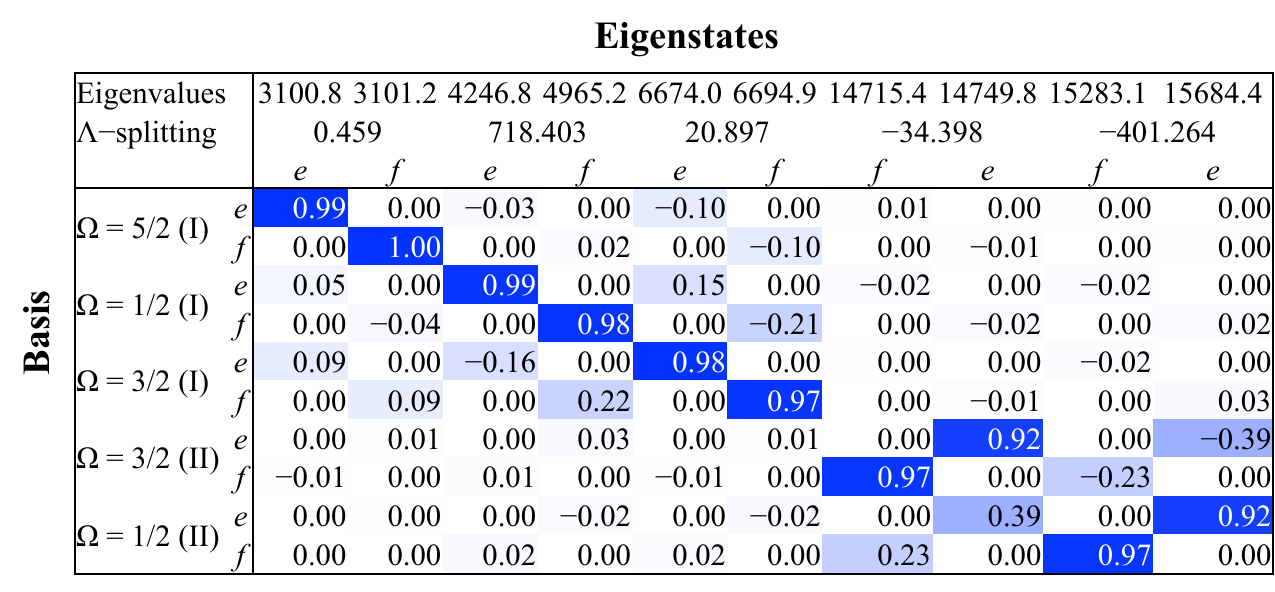}
    \caption{Normalized coefficient matrix of PtH in the $J = 20.5$ state obtained from the diagonalization of the Hamiltonian Eq.~\ref{eq:ele_rot_H}. The column is the $e/f$ basis defined in Eq.~\ref{eq:e_f_basis} for each $\Omega$ listed in Table~\ref{tbl:Energy_PtH}, The row is the eigenstates obtained from the diagonalization, where the eigenvalues and $\Lambda$-splitting are listed in \cm. The input parameters for the TW-B shown in  Table~\ref{tbl:splitting_PtH} are employed, i.e., $B_0$ = 7.177~48 \cm\ and the calculated energy~\cite{Fleig1996JMS}. The deviation between the electronic excitation energies (cf. Table~\ref{tbl:Energy_PtH}) and the provided eigenvalues in this figure arises from the rotational energy, e.g., the term $B_0J(J+1)$.}
    \label{fig:coefficients}
\end{figure*}

The coefficient matrix of PtH at the $J = 20.5$ rotational state is visualized in FIG.~\ref{fig:coefficients}. $\Omega=5/2$ basis dominantly contributes to the electronic ground $e/f$ states, but a strong mixing is observed between the $\Omega=1/2$~(II) and $3/2$~(II) states. Note that the eigenvalues and diagonal parts of the matrix elements are far from the electronic energies shown in Table~\ref{tbl:Energy_PtH} because of the contribution from the first line of \Eq{eq:rot_ham_expand} which are the rotational energies for the states in question.

Overall, our approach yields results of similar or even better quality than those obtained in Ref.~\citenum{Fleig1996JMS} where vibrational overlap has been taken into account.

%
%

%
%
\begin{table*}
\caption{
$\Lambda$-splitting ($E_f-E_e$) of PtH in \cm\ calculated with various excitation energies shown in Table \ref{tbl:Energy_PtH} and experimental rotational constants $B_0$ for the $\Omega=3/2$ (I) state (7.177~48 \cm, noted as I), and $\Omega=3/2$ (II) state (6.821~63 \cm, noted as II)~\cite{Mccarthy1993JMS}. 
PW and TW-X (X = A to E) refer to the calculated values in previous work~\cite{Fleig1996JMS}, and in this work, respectively. $\Delta$ is the mean absolute percentage error from the experiment in percentage.
}\label{tbl:splitting_PtH}
\begin{tabular}{@{}rr rr rr rr rr@{}}
\hline\hline
   & PW\cite{Fleig1996JMS} &  TW-A  &   TW-B  &  TW-C  &   TW-D$\:$  &  TW-E$\quad$   & exp.~\cite{Mccarthy1993JMS} \tabularnewline 
$B_0$ &  & (I)$\quad$ & (I)$\quad$ & (I)$\quad$ & (I)$\quad$ & (II)$\quad$ & \tabularnewline
Energy &  & \citenum{Fleig1996JMS}\footnote{Only the three lowest states ($\Omega=5/2$, $\Omega=1/2$ (I), and $\Omega=3/2$ (I)) are used in the matrix representation.  } $\quad$ & \citenum{Fleig1996JMS}$\quad$ & \citenum{Irikura2023JCP}$\quad$ & exp.~\cite{Mccarthy1993JMS}\footnote{The theoretical excitation energies of Ref.~\cite{Irikura2023JCP}\label{fn:1/2} are employed for the $\Omega=1/2$ states.} & exp.~\cite{Mccarthy1993JMS}\textsuperscript{\ref{fn:1/2}} &  \tabularnewline
\hline
$J$ &\multicolumn{3}{l}{\quad\: 0 \cm \;($\Omega=5/2$)} &  &  &  &   \tabularnewline
2.5 & 0.000 & 1$\times10^{-5}$ & 1$\times10^{-5}$ & 6$\times10^{-6}$ & 8$\times10^{-6}$ & 6$\times10^{-6}$ & \tabularnewline
3.5 & 0.000 & 7$\times10^{-5}$ & 8$\times10^{-5}$ & 4$\times10^{-5}$ & 5$\times10^{-5}$ & 4$\times10^{-5}$ & \tabularnewline
4.5 & 0.000 & 2$\times10^{-4}$ & 3$\times10^{-4}$ & 1$\times10^{-4}$ & 2$\times10^{-4}$ & 1$\times10^{-4}$ & \tabularnewline
5.5 & 0.000 & 6$\times10^{-4}$ & 7$\times10^{-4}$ & 4$\times10^{-4}$ & 5$\times10^{-4}$ & 4$\times10^{-4}$ & \tabularnewline
10.5 & 0.008 & 0.015 & 0.017 & 0.008 & 0.011 & 0.008 & \tabularnewline
15.5 & 0.053 & 0.099 & 0.114 & 0.055 & 0.071 & 0.055 & \tabularnewline
20.5 & 0.213 & 0.396 & 0.459 & 0.218 & 0.280 & 0.218 & \tabularnewline
 &  &  &  &  &  &  & \tabularnewline
$J$&\multicolumn{4}{l}{\quad 2014.4 \cm \;($\Omega=1/2$ (I))}  &  &  &   \tabularnewline
0.5 & 27.506 & 35.324 & 35.324 & 33.573 & 35.324 & 33.573 & \tabularnewline
1.5 & 55.000 & 70.630 & 70.631 & 67.107 & 70.604 & 67.107 & \tabularnewline
2.5 & 82.469 & 105.899 & 105.905 & 100.566 & 105.794 & 100.566 & \tabularnewline
3.5 & 109.898 & 141.114 & 141.128 & 133.913 & 140.853 & 133.911 & \tabularnewline
4.5 & 137.277 & 176.257 & 176.285 & 167.112 & 175.740 & 167.108 & \tabularnewline
5.5 & 164.592 & 211.310 & 211.360 & 200.129 & 210.414 & 200.123 & \tabularnewline
10.5 & 299.757 & 384.652 & 384.965 & 361.459 & 379.436 & 361.427 & \tabularnewline
15.5 & 431.375 & 553.452 & 554.413 & 514.628 & 539.148 & 514.540 & \tabularnewline
20.5 & 557.753 & 716.269 & 718.403 & 658.556 & 688.595 & 658.389 & \tabularnewline
 &  &  &  &  &  &  & \tabularnewline
$J$& \multicolumn{4}{l}{\quad 3227.7 \cm \;($\Omega=3/2$ (I))}  &  &  &   \tabularnewline
1.5 & 0.010 & 0.018 & 0.016 & 0.037 & 0.043 & 0.037 & 0.028\tabularnewline
2.5 & 0.041 & 0.073 & 0.062 & 0.147 & 0.172 & 0.147 & 0.106\tabularnewline
3.5 & 0.104 & 0.182 & 0.155 & 0.365 & 0.428 & 0.367 & 0.262\tabularnewline
4.5 & 0.212 & 0.364 & 0.308 & 0.726 & 0.851 & 0.730 & 0.518\tabularnewline
5.5 & 0.376 & 0.634 & 0.538 & 1.263 & 1.479 & 1.269 & 0.901\tabularnewline
10.5 & 2.868 & 3.899 & 3.292 & 7.549 & 8.793 & 7.586 & 5.500\tabularnewline
15.5 & 17.882 & 11.636 & 9.757 & 21.603 & 24.958 & 21.703 & 16.353\tabularnewline
20.5 &  & 25.142 & 20.897 & 44.260 & 50.639 & 44.455 & 35.079\tabularnewline
$\Delta\;(\%)$  & (51) & (30) & (41) & (36) & (58) & (36) &  \tabularnewline
 &  &  &  &  &  &  & \tabularnewline
$J$&\multicolumn{4}{l}{\quad 11247.3 \cm \;($\Omega=3/2$ (II))}   &   &  & \tabularnewline
1.5 & --0.063 &  & --0.029 & --0.018 & --0.079 & --0.068 & --0.035\tabularnewline
2.5 & --0.158 &  & --0.115 & --0.070 & --0.312 & --0.270 & --0.130\tabularnewline
3.5 & --0.316 &  & --0.287 & --0.175 & --0.772 & --0.669 & --0.315\tabularnewline
4.5 & --0.553 &  & --0.570 & --0.349 & --1.524 & --1.323 & --0.621\tabularnewline
5.5 & --3.509 &  & --0.992 & --0.607 & --2.625 & --2.281 & --1.071\tabularnewline
10.5 & --10.971 &  & --5.923 & --3.683 & --14.584 & --12.813 & --6.200\tabularnewline
15.5 & --25.100 &  & --16.893 & --10.754 & --37.569 & --33.476 & --17.064\tabularnewline
20.5 &  &  & --34.398 & --22.553 & --68.635 & --61.985 & --33.553\tabularnewline
$\Delta\;(\%)$ & (66) &  & (8) & (42) & (133) & (104) &  \tabularnewline
 &  &  &  &  &  &  & \tabularnewline
$J$&\multicolumn{4}{l}{\quad 11931.7 \cm \;($\Omega=1/2$ (II))}   &  &  & \tabularnewline
0.5 & --13.167 &  & --20.843 & --19.810 & --20.843 & --19.810 & \tabularnewline
1.5 & --26.318 &  & --41.656 & --39.601 & --41.607 & --39.551 & \tabularnewline
2.5 & --33.439 &  & --62.409 & --59.354 & --62.212 & --59.154 & \tabularnewline
3.5 & --52.514 &  & --83.073 & --79.052 & --82.586 & --78.558 & \tabularnewline
4.5 & --65.529 &  & --103.620 & --98.675 & --102.663 & --97.701 & \tabularnewline
5.5 & --78.467 &  & --124.022 & --118.207 & --122.383 & --116.534 & \tabularnewline
10.5 & --141.459 &  & --223.062 & --213.942 & --214.366 & --204.816 & \tabularnewline
15.5 & --200.108 &  & --315.698 & --305.327 & --294.915 & --282.616 & \tabularnewline
20.5 & --252.309 &  & --401.264 & --391.461 & --366.782 & --352.056 & \tabularnewline
\hline\hline
\end{tabular}
\end{table*}

%
%
\subsection{Thorium fluoride cation}\label{SUBSEC:ThF+}
\subsubsection{Kramers-restricted configuration interaction (KRCI) computation and input parameters}\label{SUBSUBSEC:ThF+_input}
The expectation values of $\braket{\hat{L}_z}$ and $\braket{\hat{S}_z}$, and the CI coefficients $C$ were calculated in this study.
The employed CI model for the lowest two electronic states is abbreviated as TZ/SD6\_CAS2in3\_SDTQ8/6a.u. and for all other states TZ/SD6\_CAS2in3\_SD8/6a.u. and comprises a triple-zeta basis set \cite{5fbasis-dyall}. Both CI models allow for single and double holes in all of the F 2p shells, a complete active space where all occupations with two electrons in the three thorium 7s and 6d$_{\delta}$ spinors and up to double (SD) excitations or up to quadruple (SDTQ) excitations into all virtual spinors below a cutoff of $6$ a.u.

For the rotational-coupling calculations in \ce{ThF+} the coupled electronic configurations are $7s^2$ and $7s7p$. $7s^2$ is the dominant configuration of the ground $^1\Sigma_0$ state, and $7s7p$ is the only configuration that can contribute to the $^1\Pi_1$ state and can couple with the $7s^2$ configuration.
The values of $d$ for the states with $S=1$ are set to 0, because we consider only the coupling between $^1\Sigma_0$ and the states coupled with it. The states represented with $^{1,3}\Pi_1$ are treated as $^{1}\Pi_1$ in the $\Lambda$-doublet calculation. The values of $d$ for all basis functions are summarized in Table~\ref{tbl:ThF+_lin_comb}.

The choice of $L$ in a multireference system can be somewhat arbitrary. In this study, we employed $L=2$ and $L=3$. The $L=2$ ($^2D_{3/2}, 6d 7s^2$) corresponds to the ground state of the Th$^+$ cation, and it is justified by comparing the electron affinity of F (27~400 \cm) and the ionization potential of Th$^{+}$ (97~600 \cm). Another option is $L=3$ ($^3F_2, 6d^2$), which is the first excited state of the $\mr{Th}^{2+}$ cation at the energy of 63.3 \cm\ from the ground state. The Mulliken population analysis of ThF$^{+}$ spinors supports of an ionic-bonding like electronic configuration, $\mr{Th}^{2+}$-F$^-$. The ground state of Th$^{2+}$, $^3H_4 \;(5f6d)$ does not significantly contribute to low-energy states of ThF$^+$. In the calculation of the matrix elements of \ce{ThF+}, we considered the basis functions with $S=0$.

%
%
\subsubsection{$\Lambda$-doublet splitting}

Table~\ref{tbl:splitting_ThF+} lists the $\Lambda$-splitting of the $^3\Delta_1$ state of \ce{ThF+}. Taking the electronic configuration of the coupled states into account (in the following denoted as ``with configuration''), the $\Lambda$-splitting is reduced to approximately half, which is a greater decrease than that predicted by the configuration coefficient $d$: $d=0.78$ for $^3\Delta_1$, and $d=0.87$ for $^1\Sigma_0$ state, which are listed in Table~\ref{tbl:ThF+_lin_comb}. When we take the electronic configuration into account, the $L=3$ value is closer to the experiment than the $L=2$ value. However, all models listed in the table can qualitatively reproduce the experimental results. This indicates that our model is reliable in predicting at least the order of magnitude of $\Lambda$-splitting for planning new experiments, even in systems like the present where a very complicated coupling mechanism makes an accurate calculation quite difficult.
\ce{ThF+}'s relatively large $\Lambda$-splitting -- given that a $^3\Delta_1$ state is concerned -- is mainly due to the very small energy difference between the ground $^3\Delta_1$ and $^1\Sigma_0$ states, amounting to only 314~\cm\ (cf. Table~\ref{tbl:ThF+_lin_comb}).
The uncertainty for the $J=1$ state obtained from the values with configuration ($L=2,3$) is ca. 50~\%. We would not provide the uncertainty arising from the other input parameters, $B_\mr{e}$ and electronic excitation energies. There is no rational justification for using $B_\mr{e}$ of another state than $^3 \Delta_1$ to determine the $\Lambda$-splitting of $^3\Delta_1$, or for employing calculated excitation energies (which are difficult to obtain accurately, cf. Table 3 of Ref.~\citenum{Denis2015NJP}), despite the experimental values being available.

%
%
\begin{table}[!htbp]
\caption{
$\Lambda$-splitting ($E_f-E_e$) of the $^3\Delta_1$ electronic first excited state of ThF$^+$ in MHz using 
experimental rotational constant $B_0=0.245$ \cm~\cite{heaven_ThF+_JCP2012}. 
}\label{tbl:splitting_ThF+}
\begin{tabular}{@{}r rrrr cc@{}}
\hline\hline
 & \multicolumn{2}{c}{w/o configuration} & \multicolumn{2}{c}{with configuration} &  & \tabularnewline
$J$ & $L=2$ & $L=3\;\;$ & $L=2$ & $L=3\;\;$ & exp.~\cite{Ng2022PRA_ThF+_omega}  & exp.~\cite{Ng_PhDthesis}\tabularnewline
\hline
1 & 5.3 & 10.7 & 2.2 & 4.5 & 5.29(5) & 10.0\tabularnewline
2 & 16.0 & 32.1 & 6.7 & 13.5 &  & \tabularnewline
3 & 32.1 & 64.2 & 13.5 & 27.0 &  & \tabularnewline
4 & 53.5 & 106.9 & 22.5 & 44.9 &  & \tabularnewline
5 & 80.2 & 160.4 & 33.7 & 67.4 &  & \tabularnewline
10 & 294.1 & 588.2 & 123.6 & 247.2 &  & \tabularnewline
15 & 641.7 & 1283.8 & 269.6 & 539.3 &  & \tabularnewline
20 & 1123.2 & 2247.4 & 471.9 & 943.8 &  & \tabularnewline
\hline\hline
\end{tabular}
\end{table}

\par
%
%
\subsection{Tantalum oxide cation}\label{SUBSEC:TaO+}
\subsubsection{Input parameters}
We employed $5d^2$ as the coupled electronic configurations, contributing to both $^3\Sigma$ states and $^3\Pi_1$ states (cf. Table~\ref{tbl:TaO+_lin_comb}). 
Another possible configuration pair would be $6s^2$ and $6s6p$. However, we ignore the contributions from these configurations because the $\Pi$ states to which the $6s6p$ configuration mainly contributes are located in a higher energy range than those listed in Table~\ref{tbl:TaO+_lin_comb}. 
A total of fourteen electronic states with $S=1,C\neq0$, and $d\neq0$ listed in Table~\ref{tbl:TaO+_lin_comb} are employed for building the coupling matrix.
Although \ce{TaO+} also exhibits multireference character similar to \ce{ThF+}, the complex with $L=3$ is a good choice here since the \ce{Ta+}, \ce{Ta^{2+}} and  \ce{Ta^{3+}} cations have $^5F_1$, $^4F_{3/2}$ and $^3F_{2}$ ground states, respectively~\cite{nist_asd}, all with $L=3$.

\begin{table*}[htbp!]
\caption{
The matrix elements of the molecular Hamiltonian defined in \Eq{eq:ele_rot_H} for \ce{TaO+} at $J=3$, including the configuration before diagonalization in the unit of $B_\mr{e}$. 
Electronic excitation energies that contribute to the diagonal elements are obtained from Ref.~\cite{Sunaga_Fleig_2022} (in \cm). 
The eight lowest-energy states that can contribute to the splitting of $^3\Delta_1$ ($S=1$, $C\neq0$, and $d\neq0$) are listed. The details of the states are summarised in Table~\ref{tbl:TaO+_lin_comb}.
}\label{tbl:mat_elem_TaO+}
\scalebox{1.0}{
\begin{tabular}{@{} cc| cc|cc| cc|cc| cc|cc| cc|cc @{}}
\hline\hline 
 &  &\multicolumn{2}{c|}{$^3\Delta_1$}  & \multicolumn{2}{c|}{$^3\Delta_3$}  & \multicolumn{2}{c|}{$^3\Sigma_0$}   & \multicolumn{2}{c|}{$^3\Sigma_1$}  & \multicolumn{2}{c|}{$^3\Phi_2$} & \multicolumn{2}{c|}{$^3\Pi_0$}  & \multicolumn{2}{c|}{$^3\Pi_0$}  & \multicolumn{2}{c}{$^3\Pi_1$} \tabularnewline
 &  & $e$ & $f$ & $e$ & $f$ & $e$ & $f$ & $e$ & $f$ & $e$ & $f$ & $e$ & $f$ & $e$ & $f$ & $e$ & $f$ \tabularnewline
 \hline
\multirow{2}{*}{$^3\Delta_1$} & $e$ & 7.003 & 0 & 0 & 0 &  --0.291 & 0 & 0 & 0 & 0 & 0 &  --0.012 & 0 &  --0.012 & 0 & 0 & 0\tabularnewline
 & $f$ & 0 & 7.003 & 0 & 0 & 0 & 0 & 0 & 0 & 0 & 0 & 0 & 0 & 0 & 0 & 0 & 0\tabularnewline
 \hline
\multirow{2}{*}{$^3\Delta_3$} & $e$ & 0 & 0 & 3184 & 0 & 0 & 0 & 0 & 0 &  --0.176 & 0 & 0 & 0 & 0 & 0 & 0 & 0\tabularnewline
 & $f$ & 0 & 0 & 0 & 3184 & 0 & 0 & 0 & 0 & 0 &  --0.176 & 0 & 0 & 0 & 0 & 0 & 0\tabularnewline
 \hline
\multirow{2}{*}{$^3\Sigma_0$} & $e$ &  --0.291 & 0 & 0 & 0 & 8935 & 0 &  --10.085 & 0 & 0 & 0 & 0 & 0 & 0 & 0 &  --10.888 & 0\tabularnewline
 & $f$ & 0 & 0 & 0 & 0 & 0 & 8935 & 0 & 0 & 0 & 0 & 0 & 0 & 0 & 0 & 0 & 0\tabularnewline
 \hline
\multirow{2}{*}{$^3\Sigma_1$} & $e$ & 0 & 0 & 0 & 0 &  --10.085 & 0 & 9381 & 0 & 0 & 0 &  --0.428 & 0 &  --0.428 & 0 & 0 & 0\tabularnewline
 & $f$ & 0 & 0 & 0 & 0 & 0 & 0 & 0 & 9381 & 0 & 0 & 0 & 0 & 0 & 0 & 0 & 0\tabularnewline
 \hline
\multirow{2}{*}{$^3\Phi_2$} & $e$ & 0 & 0 &  --0.176 & 0 & 0 & 0 & 0 & 0 & 17968 & 0 & 0 & 0 & 0 & 0 & 0 & 0\tabularnewline
 & $f$ & 0 & 0 & 0 &  --0.176 & 0 & 0 & 0 & 0 & 0 & 17968 & 0 & 0 & 0 & 0 & 0 & 0\tabularnewline
 \hline
\multirow{2}{*}{$^3\Pi_0$} & $e$ &  --0.012 & 0 & 0 & 0 & 0 & 0 &  --0.428 & 0 & 0 & 0 & 18683 & 0 & 0 & 0 &  --0.462 & 0\tabularnewline
 & $f$ & 0 & 0 & 0 & 0 & 0 & 0 & 0 & 0 & 0 & 0 & 0 & 18683 & 0 & 0 & 0 & 0\tabularnewline
 \hline
\multirow{2}{*}{$^3\Pi_0$} & $e$ &  --0.012 & 0 & 0 & 0 & 0 & 0 &  --0.428 & 0 & 0 & 0 & 0 & 0 & 18854 & 0 &  --0.462 & 0\tabularnewline
 & $f$ & 0 & 0 & 0 & 0 & 0 & 0 & 0 & 0 & 0 & 0 & 0 & 0 & 0 & 18854 & 0 & 0\tabularnewline
 \hline
\multirow{2}{*}{$^3\Pi_1$} & $e$ & 0 & 0 & 0 & 0 &  --10.888 & 0 & 0 & 0 & 0 & 0 & --0.462 & 0 & --0.462 & 0 & 18946 & 0\tabularnewline
 & $f$ & 0 & 0 & 0 & 0 & 0 & 0 & 0 & 0 & 0 & 0 & 0 & 0 & 0 & 0 & 0 & 18946\tabularnewline
\hline
\hline 
\end{tabular}
}
\end{table*}

\subsubsection{$\Lambda$-doublet splitting}
The coupling matrix in units of $B_\mr{e}$ is presented in Table~\ref{tbl:mat_elem_TaO+}. 
The magnitude of the off-diagonal elements depends on the coefficients $C$ and $d$ defined in Eq.~(\ref{eq:d_appear}). For example, the ratio between $\braket{^3\Delta_1|\hat{H}^{\mr{ROT}}|^3\Sigma_0}=-0.291$ and $\braket{^3\Pi_1|\hat{H}^{\mr{ROT}}|^3\Sigma_0}=-10.888$ is about 0.0267. This ratio is similar in amount to $C=0.0283$ for the $^3\Pi_1$ basis function of the $^3\Delta_1$ ground state presented in Table~\ref{tbl:TaO+_lin_comb}. The small discrepancy is due to a small contribution from the $^3\Sigma_1$ basis function to the $^3\Pi_1$ state, that is, the $^3\Pi_1$ state does not purely consist of the $^3\Pi_1$ basis function. 
The employed values of $d$ for the $^3\Pi_1$ state with the excitation energy of 2.348 eV and $^3\Pi_1$ correction basis function in the ground $^3\Delta_1$ state are identical. 
Although a tiny coupling between $^3\Delta_1$ and $^3\Pi_0$ is also observed due to the correction term ($^3\Pi_1$ and $^3\Sigma_0$ basis functions, respectively), the coupling that dominantly causes the $\Lambda$-splitting of $^3\Delta_1$ state is the $^3\Delta_1$ and $^3\Sigma_0$ states, through the matrix elements of the L-uncoupling term between $^3\Pi_1$ and $^3\Sigma_0$ the analytical expression of which is shown in \Eq{eq:3Sigm0_3Pi1}. 

The $\Lambda$-splitting of the $^3\Delta_1$ state of \ce{TaO+} are listed in Table~\ref{tbl:splitting_TaO+}. 
As to be seen, the employed electronic excitation energy does not affect the order of magnitude of the $\Lambda$-splitting. The uncertainty for the $J=1$ state obtained from the values in the rightmost column and in the third column from the right
is ca. 10~\%.
The $\Lambda$-splitting is much smaller than in ThF$^+$ (--4.5 MHz, shown in Table~\ref{tbl:splitting_ThF+}). The reason for this is that in TaO$^+$ (i) the energy difference between the states responsible for the $\Lambda$-splitting ($^3\Delta_1$ and $^3\Sigma_0$) is much larger 
(ii) the contribution from the $^3\Pi_1$ state to the $^3\Delta_1$ state is smaller, and (iii) the contributions from the $5d^2$ configurations to the corresponding states are smaller (cf. Table~\ref{tbl:TaO+_lin_comb}).

%
%
\begin{table}
\caption{
$\Lambda$-splitting ($E_f-E_e$) of the $^3\Delta_1$ state of TaO$^+$ in kHz, taking the configuration into account. 
The calculated vertical excitation energies ($T_\mr{e}$) reported in Refs.~\citenum{PhysRevA.95.022504} (for the six lowest-energy states up to $^3\Sigma_1$) and \citenum{Sunaga_Fleig_2022} (for the other states) are employed. Calculated rotational constant $B_\mr{e}=0.410$~\cm~\cite{PhysRevA.95.022504} or $B_\mr{e}=0.401$~\cm~\cite{Sunaga_Fleig_2022} is employed.
}\label{tbl:splitting_TaO+}
\begin{tabular}{@{} rrrr @{}}
\hline\hline
 $B_\mr{e}$ & \citenum{Sunaga_Fleig_2022}  &\citenum{PhysRevA.95.022504}  & \citenum{PhysRevA.95.022504}\tabularnewline
$T_\mr{e}$  &  \citenum{Sunaga_Fleig_2022} &  \citenum{Sunaga_Fleig_2022}   &(\citenum{PhysRevA.95.022504} and \citenum{Sunaga_Fleig_2022})\tabularnewline
 \hline
$J\;$ 1 &  7.9 & 8.2 & 8.7\tabularnewline
2 &  23.6 & 24.6 & 26.1\tabularnewline
3 &  47.1 & 49.2 & 52.1\tabularnewline
4 &  78.5 & 82.0 & 86.9\tabularnewline
5 &  117.8 & 123.0 & 130.3\tabularnewline
10 &  432.0 & 451.0 & 477.7\tabularnewline
15 &  942.6 & 984.0 & 1042.4\tabularnewline
20 & 1649.6 & 1721.9 & 1824.2\tabularnewline
\hline\hline
\end{tabular}
\end{table}

\subsection{Conditions for $\Lambda$-splitting}\label{SEC:SMALL_LAMBDA}
Molecules with small $\Lambda$-splitting are suitable for CP-violation search, since small external electric fields lead to full polarization. 
For diatomic molecules with a low-lying multiplet of $\Lambda\text{-}S$ states such as $^{1,3}\Delta$, $^{1,3}\Sigma$, and $^{1,3}\Pi$, we can identify several decisive characteristics.
One factor for achieving small $\Lambda$-splitting is the energy difference between the $^3\Delta_1$ (or $\Sigma_1$ and $\Pi_1$ in case they are the target states)
and $\Sigma_0$ states, because the $\Lambda$-splitting decreases as the energy difference increases in perturbative expressions \cite{brion,Fan2025PRA_NEQ_doublet}.
Another factor is the strength of the spin-orbit interaction. The $\Lambda$-splitting of the $^3\Delta_1$ state becomes zero when there is no $\Pi_1$ contribution to the $^3\Delta_1$ state. This contribution is driven by the magnitude of SO coupling, a relativistic effect.
From this point of view intermediately heavy molecular systems, e.g., \ce{HfF+} and \ce{TaO+}, have an advantage over heavier systems such as ThO and \ce{ThF+},
although the nuclear charge is not the only factor that determines the size of the SO coupling. For instance, Table~\ref{tbl:Energy_PtH} shows larger mixing between $^{2}\Pi_{3/2}$ and $^{2}\Delta_{3/2}$ basis functions of PtH than the corresponding mixing in the case of \ce{ThF+}. 
Finally, in the case of multi-reference systems, the rotational coupling between $\Pi_1$ and $\Sigma_0$ basis functions is reduced when the electronic configurational overlap between them is small.

\section{Conclusion}
\label{SEC:CONCL}
We present a theoretical model and calculations of the rotational-coupling effects in three molecular species, PtH, \ce{ThF+}, and \ce{TaO+}. 
Our model integrates the multi-reference four-component wavefunction and the $\Lambda$-splitting based on Hund's case (a). 
The matrix elements for the rotational-coupling Hamiltonian are constructed employing the electronic excitation energies and rotational constants as input parameters, and are diagonalized in an $e$- and $f$-type basis.
In our model, the $\Lambda$-splitting of the $^3\Delta_1$ state occurs indirectly through the $J^{\pm}L^{\mp}$ operators acting on $\Pi_1$ and $\Sigma_0$ states. Thus, it is an $L$-uncoupling effect entering through higher orders in perturbation theory.
The dominant factors that determine the magnitude of the splitting of the target $^3\Delta_1$ state are (i) the contribution of $\Pi_1$ to the $^3\Delta_1$ state (i.e., the strength of the spin-orbit coupling), (ii) the energy difference between the $^3\Delta_1$ state and the $\Sigma_0$ state, (iii) the electronic configurational overlap of the $\Sigma_0$ and the $\Pi_1$ basis functions. Our calculations of the $\Lambda$-splitting qualitatively agree with the available experimental data for PtH and \ce{ThF+}.

The $\Lambda$-splitting of \ce{TaO+} is in our present work predicted to be about $9$ kHz in the electronic ground state $^3\Delta_1$ and for rotational quantum number $J=1$.
This small $\Lambda$- ($\Omega$-) splitting can reduce the systematic uncertainty due to the external electric field. In addition, 
this is small enough to avoid the use of larger ring traps and the study of ion dynamics in the ring trap. On the other hand it may not be large enough to avoid depolarization during rotation ramp-up \cite{Zhou_private}.

Our program is also applicable for estimating the $\Lambda$-doubling of other target molecules for CP-violation search, such as TaN ($^3\Delta_1$)~\cite{Skripnikov2015PRA_TaN,Fleig2016PRA}, WC ($^3\Delta_1$)~\cite{Lee2009JMO_WC}, and PbO ($^3\Sigma_1$)~\cite{DeMille2000PRA_PbO,Kozlov2002PhysPRL_PbO,Eckel2013PRA_PbO} molecules.
$\Omega$-doublings from a Hund's case (c) point of view~\cite{brion,Veseth1973JPB_Hund_c_I,Veseth1973JPB_Hund_c_II} could be obtained with only minor modifications of the present formulation.

\begin{acknowledgments}
We thank Dr. Yan Zhou (Las Vegas) for helpful discussions. A.S. acknowledges financial support from the Japan Society for the Promotion of Science (JSPS) KAKENHI (Grant No. 21K14643).
\end{acknowledgments}

%
%
\appendix
\section{Relative sign of the linear combination and eigenvalues}\label{app:sign}

Consider the following two matrices:
\be
\bos{A}= \left(\begin{array}{cccc}E_{a} & 0 & V_{ab} & 0 \\ 0 & E_{a} & 0 & 0 \\ V_{ab} & 0 & E_{b} & 0 \\ 0 & 0 & 0& E_{b} \end{array}\right)
;\quad 
\bos{P}= \left(\begin{array}{cccc}1 & 0 & 0 & 0 \\ 0 & 1 & 0 & 0 \\ 0 & 0 & -1 & 0 \\ 0 & 0 & 0& 1 \end{array}\right)
\ee
where $\bos{P}$ is an invertible matrix.
Assuming $\bos{A}$ is a part of the matrix 
provided in Table~\ref{tbl:mat_elem_TaO+} (e.g., $e/f$ of $^3\Delta_1$ and $e/f$ of $^3\Sigma_0$), the sign of $V_{ab}$ depends on the relative sign of the linear combination defined in \Eq{eq:lin_comb}. Next, we define a matrix similar to a $\bos{A}$ as follows:
\be
\bos{A}'= \bos{P}^{-1}\bos{A}\bos{P} = \left(\begin{array}{cccc}E_{a} & 0 & -V_{ab} & 0 \\ 0 & E_{a} & 0 & 0 \\ -V_{ab} & 0 & E_{b} & 0 \\ 0 & 0 & 0& E_{b} \end{array}\right)
\ee
Since $\bos{A}$ and $\bos{A}'$ are similar their eigenvalues are identical. 
This similarity transformation 
shows that the size of the $\Lambda$-splitting and the ordering of the $e$ and $f$ states do not depend on the relative sign of the linear combination when the coupling between the two states is dominant.
This condition is satisfied in our case: for \ce{ThF+}, the coupling between $^3\Delta_1$ and $^1\Sigma_0$ is dominant because of their tiny energy difference (314~\cm, cf. Table \ref{tbl:ThF+_lin_comb}), and for \ce{TaO+}, the dominant coupling is between $^3\Delta_1$ and $^3\Sigma_0$ (cf. Table \ref{tbl:mat_elem_TaO+}).
This indicates that our model, which ignores the relative sign, works well. However, if several states couple to the target state with comparable strength, a partial change in the sign of the matrix elements due to the relative signs of the linear combination could lead to a sign dependence of the eigenvalue. 

\section{Electronic structure data}\label{app:data}

Tables~\ref{tbl:Energy_PtH}-\ref{tbl:TaO+_lin_comb} list the employed excitation energies, CI coefficients ($C$), and configuration parameters ($d$, defined in \Eq{eq:d_appear}) of the PtH, \ce{ThF+}, and \ce{TaO+} molecules, respectively.  Table~\ref{tbl:ThF+_lin_comb} (\ref{tbl:TaO+_lin_comb}) lists both the singlet and triplet states, but only the singlet (triplet) basis functions are employed for the calculation of Table~\ref{tbl:splitting_ThF+} (\ref{tbl:splitting_TaO+}). Although our previous work~\cite{Sunaga_Fleig_2022} provided higher-energy electronic excited states, Table~\ref{tbl:TaO+_lin_comb} presents only the states to which $6s$ and $5d$ orbitals dominantly contribute. 

%
%
\begin{table}[htbp!]
\caption{
Calculated and experimental excitation energy of PtH. 
The values of Ref.~\citenum{Fleig1996JMS} are the vertical excitation energies (in~\cm), while those of Refs~\citenum{Irikura2023JCP} and~\citenum{Mccarthy1993JMS} are the energies of the vibrational ground states. 
The coefficients of the linear combinations of the $\ket{\Lambda_{\Omega}}$ basis correspond to $C$ defined in \Eq{eq:e_f_lincomb}. All states have $S=1/2$.
}\label{tbl:Energy_PtH}
\begin{tabular}{@{}l rrr@{}}
\hline \hline 
 &\multicolumn{2}{c}{Theory} & Experiment  \tabularnewline
\quad\quad\quad\quad\quad\quad\citenum{Fleig1996JMS}$^\mr{a}$ & \citenum{Fleig1996JMS}\footnote{Taken from $e$ states listed in Table 5.} & \citenum{Irikura2023JCP}\footnote{Taken from the zero-point energies shown in a file v-dependent\_constants\_PtH.xlsx in the Supplementary Material.}  & \citenum{Mccarthy1993JMS}\footnote{Taken from Table VII of Ref.~\citenum{Mccarthy1993JMS}.} \tabularnewline
\hline 
$\Delta_{5/2}$ & 0 & 0 & 0\tabularnewline
$-0.939\ket{\Sigma_{1/2}}-0.343\ket{\Pi_{1/2}}$ & 1479.2 & 2014.4 & \tabularnewline
$\;\;\; 0.766\ket{\Pi_{3/2}}+0.643\ket{\Delta_{3/2}}$  & 3414.2 & 3227.7 & 3224.89\tabularnewline
$-0.767\ket{\Delta_{3/2}}-0.641\ket{\Pi_{3/2}}$  & 11625.9 & 11247.3 & 11581.55\tabularnewline
$\;\;\; 0.940\ket{\Pi_{1/2}}-0.340\ket{\Sigma_{1/2}}$  & 12208.4 & 11931.7 & \tabularnewline
\hline\hline 
\end{tabular}
\end{table}

\begin{table*}[htbp!]
\caption{Hund's case (a) basis and coefficients for ThF$^+$. The singlet basis functions are considered for the computation of Table~\ref{tbl:splitting_ThF+}.
The coefficients for the linear combination are obtained from the values of $\braket{\hat{S}_z}$ and $\braket{\hat{L}_z}$ of each state obtained with the KRCI method (see Sec.~\ref{SUBSUBSEC:ThF+_input}). For example, the electronic first excited state $\ket{\Psi_e}$ can be expressed by $\ket{\Psi_e}= 0.9911\ket{{ }^3 \Delta_1} + 0.1330\ket{{ }^1 \Pi_1}$. The vertical excitation energies $T_\mr{e}$ of the states are in~\cm.
The spectroscopic term of the dominant basis state was assigned referring to Table 9 of Ref.~\cite{Denis2015NJP}. 
The molecular spinors that do not provide $\lambda$ indicate that $\lambda$ is not a good quantum number.
}\label{tbl:ThF+_lin_comb}
\begin{tabular}{@{} r rr cc cc cc ll @{}}
\hline\hline
 & \multicolumn{2}{c}{KRCI} & \multicolumn{3}{c}{dominant} & \multicolumn{3}{c}{correction}  & \multicolumn{2}{c}{configuration}\tabularnewline
$T_\mr{e}$ & $\braket{\hat{S}_z}\;\;$ & $\braket{\hat{L}_z}\;\;$  & basis & $C$ & $d$ & basis & $C$ & $d$ & $C_\mr{CI}^2$ & Th$\left(n l_{\lambda, \omega}\right)$ \tabularnewline

\hline
0 & $-$0.9822 & 1.9822 & $^3\Delta_1$ & 0.9911 & 0 & $^1\Pi_1$ & 0.1334 & 0.78 & 0.94 & $ 7s_{\sigma,1/2}, 6d_{\delta,3/2} $ \tabularnewline
\hline
314\footnote{Experimental data~\cite{Gresh_ThF+_JMS2016}} & 0.0000 & 0.0000 & $^1\Sigma_0$ & 1.0000 & 0.87  & -- & 0.0000   & 0 & 0.75 & $ (7s_{\sigma,1/2})^2 $  \tabularnewline
&& &  &  &  &  &  &  & 0.12 & $ (6d_{\delta,3/2})^2 $  \tabularnewline
\hline
3395\footnote{Experimental data~\cite{heaven_ThF+_JCP2012}} & $-$0.0100 & 0.0100 & $^1\Sigma_0$ & 0.9950 & 0 & $^3\Pi_0$ & 0.1000 & 0 & 0.58 & $ (6d_{\delta,3/2})^2 $ \tabularnewline
&& &  &  &  &  &  &  & 0.12 & $ (6d_{\delta,5/2})^2 $ \tabularnewline
\hline
6528\footnote{Calculated data (Table 9 of Ref.~\cite{Denis2015NJP})} & $-$0.9471 & 0.9471 & $^3\Pi_0$ & 0.9732 & 0 & $^1\Sigma_0$ & 0.2300 & 0.87 & 0.29 & $  7s_{\sigma,1/2} , 7p6d_{1/2}  $ \tabularnewline
&& &  &  &  &  &  &  & 0.29 & $ 7s_{\sigma,1/2} ,  7p6d_{1/2} $  \tabularnewline
&& &  &  &  &  &  &   & 0.11 & $  7s_{\sigma,1/2}, 7p6d_{1/2}  $  \tabularnewline
&& &  &  &  &  &  &   & 0.11 & $ 7s_{\sigma,1/2} ,  7p6d_{1/2} $  \tabularnewline
\hline
6639$^{\text{c}}$ & 0.1092 & 0.8908 & $^{1,3}\Pi_1$ & 0.9438 & 0.78 & $^3\Sigma_1$ & 0.3305 & 0 & 0.41 & $ 7s_{\sigma,1/2} ,7p6d_{1/2} $ \tabularnewline
&& &  &  &  &  &  &  & 0.18 & $ 7s_{\sigma,1/2} ,7p6d_{1/2} $  \tabularnewline
&& &  &  &  &  &  &   & 0.16 & $ 6d_{\delta,3/2}, 6d_{\delta,5/2} $  \tabularnewline
\hline
6747$^{\text{c}}$ & 0.9485 & $-$0.9485 & $^3\Pi_0$ & 0.9739 & 0 & $^1\Sigma_0$ & 0.2269 & 0.87 & 0.23 & $ 7s_{\sigma,1/2} , 7p6d_{1/2}  $ \tabularnewline
&& &  &  &  &  &  &  & 0.23 & $ 7s_{\sigma,1/2} ,  7p6d_{1/2} $  \tabularnewline
&& &  &  &  &  &  &  & 0.11 & $ 6d_{\delta,3/2}  , 6d_{\delta,3/2} $ \tabularnewline
&& &  &  &  &  &  &   & 0.09 & $ 7s_{\sigma,1/2}, 7p6d_{1/2} $  \tabularnewline
&& &  &  &  &  &  &   & 0.09 & $ 7s_{\sigma,1/2} ,  7p6d_{1/2} $  \tabularnewline
\hline
7490$^{\text{c}}$ & 0.9890 & 0.0110 & $^{3}\Sigma_1$ & 0.9945 & 0 & $^1\Pi_1$ & 0.1049 & 0.78 & 0.74 & $ 6d_{\delta,3/2} ,6d_{\delta,5/2} $ \tabularnewline
\hline
7918$^{\text{c}}$ & 0.0003 & 0.9997 & $^{1,3}\Pi_1$ & 0.9998 & 0.20 & $^{3}\Sigma_1$ & 0.0173 & 0 & 0.42 & $ 7s_{\sigma,1/2}, 6d7p_{3/2} $ \tabularnewline
&& &  &  &  &  &  & & 0.19 & $ 7p6d_{1/2}, 6d_{\delta,3/2} $  \tabularnewline
&& &  &  &  &  &  &   & 0.15 & $ 7s_{\sigma,1/2} ,7p6d_{3/2} $  \tabularnewline
\hline\hline
 
\end{tabular}
\end{table*}

%
%
\begin{table*}[htbp!]
\caption{Hund's case (a) basis and coefficients for TaO$^+$. 
The triplet basis functions are considered for the computation of Table~\ref{tbl:splitting_TaO+}.
The values of $\braket{\hat{S}_z}$, $\braket{\hat{L}_z}$, and $C$ are obtained in Ref.~\cite{Sunaga_Fleig_2022}.
For example, the electronic ground state $\ket{\Psi_g}$ can be expressed by $\ket{\Psi_g}= 0.9996\ket{{ }^3 \Delta_1} + 0.0283\ket{{ }^3 \Pi_1}$. The vertical excitation energies $T_\mr{e}$ of the states are in eV. The electronic configuration shown in Ref.~\citenum{Sunaga_Fleig_2022} was corrected.
}\label{tbl:TaO+_lin_comb}
\begin{tabular}{@{}cc rr cc cc cc ll @{}}
\hline\hline
\multicolumn{2}{c}{$T_\mr{e}$ (eV)}& \multicolumn{2}{c}{KRCI} & \multicolumn{3}{c}{dominant} & \multicolumn{3}{c}{correction}  & \multicolumn{2}{c}{configuration} \\
Ref.~\citenum{PhysRevA.95.022504} & Ref.~\citenum{Sunaga_Fleig_2022}& $\braket{\hat{S}_z}\;\;$ & $\braket{\hat{L}_z}\;\;$ & basis & $C$ & $d$ & basis & $C$ & $d$  & $C_\mr{CI}^2$ & Ta$\left(n l_{\lambda, \omega}\right)$ \\ 

\hline 
0.000 & 0.000 & --0.9992 & 1.9992 & ${ }^3 \Delta_1$ & 0.9996 & 0 & ${ }^3 \Pi_1$ & 0.0283 & 0.43 & 0.89 & $6s_{\sigma,1/2}, 5d_{\delta,3/2}$\\
\hline
0.163 &0.158 & 0.0000 & 2.0000 & ${ }^3 \Delta_2$ & 1.0000 & 0 & - & - & 0 & 0.60 & $6s_{\sigma,1/2}, 5d_{\delta,3/2}$\\
&&& &&&&&&& 0.29 & $6s_{\sigma,1/2},5d_{\delta,5/2}$\\
\hline
0.405 &0.395 & 0.9974 & 2.0026 & ${ }^3 \Delta_3$ & 0.9987 & 0 & ${ }^3 \Phi_3$ & 0.0510 & 1.00 & 0.89 & $6s_{\sigma,1/2} 5d_{\delta,5/2}$\\
\hline
0.466 &0.414 & 0.0000 & 0.0000 & ${ }^1 \Sigma_0$ & 1.0000 & 0 & - & - & 0 & 0.62 & $(6s_{\sigma,1/2})^2$\\
&&& &&&&&&& 0.22 & $(5d_{\delta,3/2})^2$\\
\hline
1.025 &1.106 & 0.0000 & 0.0000 & ${ }^3 \Sigma_0$ & 1.0000 & 1.00 & - & - & 0 & 0.50 & $(5d_{\delta,3/2})^2$\\
&&& &&&&&&& 0.30 & $(5d_{\delta,5/2})^2$\\
\hline
1.043 &1.162 & 0.9992 & 0.0008 & ${ }^3 \Sigma_1$ & 0.9996 & 1.00 & ${ }^3 \Pi_1$ & 0.0283 & 0.43 & 0.88 & $5d_{\delta,3/2},5d_{\delta,5/2}$\\
\hline
1.421 &1.342 & 0.0006 & 1.9994 & ${ }^1 \Delta_2$ & 0.9997 & 0 & ${ }^3 \Pi_2$ & 0.0245 & 0 & 0.57 & $6s_{\sigma,1/2},5d_{\delta,5/2}$\\
&&& &&&&&&& 0.26 & $6s_{\sigma,1/2},5d_{\delta,3/2}$\\
\hline
&1.804 & 0.0043 & 3.9957 & ${ }^1 \Gamma_4$ & 0.9978 & 0 & ${ }^3 \Phi_4$ & 0.0656 & 1.00 & 0.88 & $5d_{\delta,5/2}, 5d_{\delta,3/2}$\\
\hline
&2.228 & --0.9965 & 2.9965 & ${ }^3 \Phi_2$ & 0.9982 & 1.00 & ${ }^3 \Delta_2$ & 0.0592 & 0 & 0.89 & $5d_{\pi,1/2}, 5d_{\delta,3/2}$\\
\hline
&2.315 & 0.9982 & --0.9982 & ${ }^3 \Pi_0$ & 0.9991 & 0 & ${ }^3 \Sigma_0$ & 0.0424 & 1.00 & 0.44 & $6s_{\sigma,1/2},5d_{\pi,1/2}$\\
&&& &&&&&&& 0.44 & $6s_{\sigma,1/2},5d_{\pi,1/2}$\\
\hline
&2.336 & --0.9982 & 0.9982 & ${ }^3 \Pi_0$ & 0.9991 & 0 & ${ }^3 \Sigma_0$ & 0.0424 & 1.00 & 0.40 & $6s_{\sigma,1/2},5d_{\pi,1/2}$\\
&&& &&&&&&& 0.40 & $6s_{\sigma,1/2},5d_{\pi,1/2}$\\
\hline
&2.348 & 0.0039 & 0.9961 & ${ }^3 \Pi_1$ & 0.9980 & 0.43 & ${ }^3 \Sigma_1$ & 0.0624 & 1.00 & 0.50 & $6s_{\sigma,1/2},5d_{\pi,1/2}$\\
&&& &&&&&&& 0.37 & $5d_{\pi,1/2},5d_{\delta,3/2}$\\
\hline
&2.417 & 0.0000 & 0.0000 & ${ }^1 \Sigma_0$ & 1.0000 & 0 & - & - & 0 & 0.47 & $(5d_{\delta,5/2})^2$\\
&&& &&&&&&& 0.13 & $(6s_{\sigma,1/2})^2$\\
&&& &&&&&&& 0.13 & $(5d_{\delta,3/2})^2$\\
\hline
&2.543 & 0.0009 & 0.9991 & ${ }^3 \Pi_1$ & 0.9995 & 0.34 & ${ }^3 \Sigma_1$ & 0.0300 & 1.00 & 0.39 & $6s_{\sigma,1/2},5d_{\pi,3/2}$\\
&&& &&&&&&& 0.30 & $5d_{\pi,1/2},5d_{\delta,3/2}$\\
&&& &&&&&&& 0.19 & $6s_{\sigma,1/2},5d_{\pi,1/2}$\\
\hline
&2.619 & 0.0053 & 2.9947 & ${ }^3 \Phi_3$ & 0.9973 & 1.00 & ${ }^3 \Delta_3$ & 0.0728 & 0 & 0.45 & $5d_{\pi,1/2},5d_{\delta,5/2}$\\
&&& &&&&&&& 0.45 & $5d_{\pi,3/2},5d_{\delta,3/2}$\\
\hline
&2.717 & 1.0000 & 1.0000 & ${ }^3 \Pi_2$ & 1.0000 & 0 & - & - & 0 & 0.84 & $6s_{\sigma,1/2},5d_{\pi,3/2}$\\
\hline
&2.880 & 0.9951 & --0.9951 & ${ }^3 \Pi_0$ & 0.9975 & 1.00 & ${ }^3 \Sigma_0$ & 0.0700 & 1.00 & 0.44 & $5d_{\pi,3/2},5d_{\delta,3/2}$\\
&&& &&&&&&& 0.44 & $5d_{\pi,3/2},5d_{\delta,3/2}$\\
\hline
&2.892 & --0.9951 & 0.9951 & ${ }^3 \Pi_0$ & 0.9975 & 1.00 & ${ }^3 \Sigma_0$ & 0.0700 & 1.00 & 0.44 & $5d_{\pi,3/2},5d_{\delta,3/2}$\\
&&& &&&&&&& 0.44 & $5d_{\pi,3/2},5d_{\delta,3/2}$\\
\hline
&2.913 & 0.9982 & 1.0018 & ${ }^3 \Pi_2$ & 0.9991 & 1.00 & ${ }^3 \Delta_2$ & 0.0424 & 0 & 0.83 & $5d_{\delta,5/2} 5d_{\pi,1/2}$\\
\hline
&3.018 & 0.9964 & 3.0036 & ${ }^3 \Phi_4$ & 0.9982 & 1.00 & ${ }^1 \Gamma_4$ & 0.0600 & 0 & 0.90 & $5d_{\delta,5/2} 5d_{\pi,3/2}$\\
\hline
&3.024 & 0.0004 & 0.9996 & ${ }^1 \Pi_1$ & 0.9998 & 0 & ${ }^3 \Sigma_1$ & 0.0200 & 1.00 & 0.45 & $5d_{\pi,3/2},5d_{\delta,5/2}$\\
&&& &&&&&&& 0.27 & $6s_{\sigma,1/2},5d_{\pi,3/2}$\\
&&& &&&&&&& 0.11 & $5d_{\pi,1/2},5d_{\delta,3/2}$\\
\hline
&3.629 & 0.0005 & 2.9995 & ${ }^1 \Phi_3$ & 0.9997 & 0 & ${ }^3 \Delta_3$ & 0.0224 & 0 & 0.43 & $5d_{\pi,1/2},5d_{\delta,5/2}$\\
&&& &&&&&&& 0.42 & $5d_{\pi,3/2},5d_{\delta,3/2}$\\
\hline\hline

\end{tabular}
\end{table*}
\clearpage

\bibliography{all,ref}
\clearpage


\end{document}